\documentclass[aps, twocolumn, prb, showpacs,preprintnumbers,amsmath,amssymb,superscriptaddress]{revtex4-1}
\usepackage{amsmath,amsfonts,amssymb,graphics,graphicx,epsfig,color,times,indentfirst,layout,hyperref,bm}
\usepackage{hyperref}
\usepackage{ulem}
\hypersetup{linktocpage,,colorlinks=true}

\newcommand{\bk}{{\bf k}}

\newcommand{\Ham}{\mathcal{H}}

\begin{document}

\title{Topology and entanglement in quench dynamics}

\author{Po-Yao Chang}
\email{pychang@physics.rutgers.edu}
\affiliation{Center for Materials Theory, Rutgers University, Piscataway, New Jersey, 08854, USA }

\begin{abstract}
We classify the topology of the quench dynamics by homotopy groups.
A relation between the topological invariant of a post-quench order parameter and
the topological invariant of a static Hamiltonian is shown in $d+1$ dimensions ($d=1,2,3$).
The mid-gap states in the entanglement spectrum of the post-quench states reveal their topological nature.
When a trivial quantum state under a sudden quench to a Chern insulator, 
the mid-gap states in the entanglement spectrum form rings.
These rings are analogous to the boundary Fermi rings in the Hopf insulators.
Finally, we show a post-quench order parameter in 3+1 dimensions can be characterized 
by the second Chern number. The number of Dirac cones in the entanglement spectrum
is equal to the second Chern number.
\end{abstract}
\maketitle

\section{Introduction}


Recently, properties of quantum states far from equilibrium attract huge attention 
due to the progressive developments in cold-atom experiments.
For example, a quantum Newton's cradle setup is realized in one-dimensional Bose gases\cite{Kinoshita2006} 
and the emergence of quantum turbulence is observed in a Bose-Einstein condensate\cite{Navon2018}.
Besides examining the dynamical properties of non-equilibrium states,
cold-atom experiments also provide a playground for engineering topological systems including
the Su-Schrieffer-Heeger model\cite{Atala2013}, the Hofstadter model\cite{Aidelsburger2013,Aidelsburger2014,Miyake2013}, and the Haldane model (Chern insulators)\cite{Jotzu2014,Wu83}.
These topological phases are typically classified by the topological invariant in the static Hamiltonian.
The classification of the topological phases out of equilibrium is one of the main interests in both condensed matter and cold-atom
communities.

To study non-equilibrium topological phases, 
one straightforward setup is Floquet systems.
The band inversion mechanism is introduced by a periodically driven source and gives rise to a Floquet topological insulator\cite{Lindner2011,Jiang2011,Gomezleon2013}.
Another setup is considering a dynamical process which an initial state of a trivial Hamiltonian
is evolved under a sudden quench to a non-trivial Hamiltonian.
This quantum quench involves the change of the topological number and can be revealed 
in the post-quench states. 
In a generic one-dimensional two-band model,
a dynamic Chern number characterizes the topological property of the post-quench state \cite{Yang2018, Gong2017}.
In a two-band Chern insulator, the dynamics of the post-quench state can be captured by the Hopf number \cite{Wang2017,Tarnowski2017}.
 The post-quench dynamics across a quantum critical point
is also influenced by the topological edge states\cite{Foster2013,Foster2014,Liao2015,Bermudez2009,Bermudez2010}.
The topology of the static Hamiltonian and topology of the post-quench state are related.

In this paper, we establish this relation in a systematic way by use of homotopy groups.
We show that the post-quench state is periodic in time under a sudden quench to a translational invariant Hamiltonian in $d$ dimensions ($d=1,2,3$).
The momentum-time space is a $d+1$ dimensional torus $T^{d+1}$.
A mapping from the time-momentum space $T^{d+1}$ to a manifold $\mathcal{M}$ of the post-quench order parameter 
is characterized by the homotopy group $\pi_{d+1}(\mathcal{M})$.
In a two-band model, the manifold of the post-quench pseudospin is a two-sphere $S^2$. 
Due to the non-vanishing homotopy groups $\pi_{1+1}(S^2)$ and $\pi_{2+1}(S^2)$,
the post-quench states have nontrivial topology in one and two dimensions.
However, in a generic $n$-band model ($n>2$), the post-quench order parameter can be a higher dimensional manifold $\mathcal{M} \neq S^{2}$.
The statement for two-band models, in general, cannot generalize to $n$-band models ($n>2$).
We consider two different strategies to overcome this obstacle.
Firstly, we project both the static Hamiltonian and the post-quench state in the sub-manifolds
that have the similar structures as a two-band model. We demonstrate this strategy by spin-1 models.
Secondly, we consider higher order homotopy groups to classify both the static Hamiltonian and the post-quench order parameter.
In a four-band model, we demonstrate the relation between the three-dimensional winding number of the static Hamiltonian [classified by $\pi_3(S^3)$]
and the second Chern number of the post-quench order parameter [classified by $\pi_4(S^4)$].

The entanglement spectrum also reveals the topological property of the post-quench states.
In 1+1 dimensional post-quench states, the entanglement spectrum has crossings when the dynamic Chern number of the post-quench order parameter is non-zero\cite{Gong2017}.
Here, we extend this analysis to 2+1 and 3+1 dimensions 
and considering two different bipartitions. 
In a real space bipartition, we show that when the post-quench state is non-trivial, mid-gap states in the entanglement spectrum
form Dirac cones in 2+1 and 3+1 dimensions. 
The number of Dirac cones in the entanglement spectrum directly links to the topological index of the post-quench states. 
We compute the topological index of the post-quench states
and show it relates to the topological invariant of the post-quench order parameters.
In a frequency space bipartition\cite{FB},
we show that the mid-gap states in the entanglement spectrum in 2+1 dimensions form rings.
In this case, the topological invariant of the post-quench order parameter is characterized by the Hopf number.
These rings in the entanglement spectrum are analogous to the boundary Fermi rings in the Hopf insulators\cite{Liu2017,Moore2008, Deng2013}. 

The rest of the paper is organized as follows: In Sec. \ref{Sec:QP},
we introduce the quench protocol and the classification scheme by homotopy groups.
In Sec. \ref{Sec:2B}, we review the quench dynamics of two-band models and 
introduce a different quench process which has not been discussed before.
In Sec. \ref{Sec:S1}, we show the quench dynamics
and the entanglement spectrum of the post-quench states in spin-1 models.
In Sec. \ref{Sec:4B}, we demonstrate that the second Chern number captures
the quench dynamics in 3+1 dimensions in a four-band model.
In Sec. \ref{Sec:CD}, we conclude our work and give some discussions.



\section{Quench protocol}\label{Sec:QP}

We consider a free-fermion Hamiltonian with discrete translational symmetries
\begin{align}
H= \sum_{\bf k} \sum_{i,j,=1}^n c^\dagger_{\bk,i} \Ham_{\bk}^{ij}c_{\bk,j},
\end{align}
where $\bk$ is the momentum, $\Ham_{\bk}^{ij}$ is the single particle Hamiltonian with $n$ being the number of bands,
and $c^\dagger_{\bk,i}$ is the fermionic operator in the momentum space.
If the single particle Hamiltonian $\Ham_{\bk}^{ij}$ has a massive Dirac Hamiltonian representation,
one can classify the possible distinctive Dirac mass term by homotopy groups\cite{Chiu2016, Kitaev2009,ABRAMOVICI2012,Morimoto2013}.
This classification requires the number of bands to be large.
On the other hand, in few-band models, the classification of the single particle Hamiltonian 
can also be obtained from the structure of the single particle Hamiltonian directly\cite{DENITTIS2014303,Liu2017,Moore2008, Deng2013,Kennedy2016,Bzdu2017}.
In particular, we can parametrize these few-band models with a finite set of functions $\{ f_{\bk, \alpha} \}$.
Each point in the momentum space (which is a $d$-dimensional torus) maps to a point in the manifold $\mathcal{M}$ of this set of functions.
We can classify distinct sets of functions by the $d$-th homotopy group $\pi_{d}(\mathcal{M})$.

The evolution operator of the many-body state is
$\hat{U}(t)= e^{-i H t}$.
Let us assume the initial many-body state $|\Psi_i\rangle$ is prepare from a given Hamiltonian $H_i$ such that 
$|\Psi_i\rangle= \prod_{\bk, \alpha \in {\rm occ. }} d^\dagger_{\bk,\alpha} |0\rangle$ with $ d_{\bk,\alpha} $ being the 
fermionic operator in the enegy basis of the initial Hamiltonian $H_i = \sum_{\bk,\alpha} \epsilon_{\bk,\alpha}   d^\dagger_{\bk,\alpha}d_{\bk,\alpha}$
and "occ." refers to the 
occupied bands.
Here $ \epsilon_{\bk,\alpha}$ is the eigen-energy of $H_i$.
The corresponding single particle wave function is $ d^\dagger_{\bk,\alpha}|0 \rangle = \sum_{i}u_{\alpha, i}c^\dagger_{\bk, i}|0\rangle$,
where 
 $u_{\alpha,i}$ is the unitary matrix diagonalizing $H_i$.

The post-quench many-body state is
\begin{align}
|\Psi(t)\rangle&=\hat{U}(t) |\Psi_i\rangle  \notag\\
&=\prod_{\bk, \alpha}e^{-i t \sum_{i,j}\Ham_{\bk}^{ij}   c^\dagger_{\bk,j} c_{\bk,j}} d^\dagger_{\bk,\alpha}    |0 \rangle \notag\\
&=\prod_{\bk, \alpha}\sum_{i,j} [e^{-i t \Ham_{\bk}}]_{ij} u_{\alpha,j}   c^\dagger_{\bk,i} |0\rangle,
\end{align}
which is factorized in the momentum $\bk$.  Hence we only need to focus on the single particle evolution operator
$U_\bk(t)=e^{-i t \Ham_k}$ acting on the single-particle wave-function.
I.e., Particles do not interchange the momentum due to its non-interacting nature .
We can consider a measure of a operator ${\bf O}_{\bk}=\sum_{i,j}c^\dagger_i \mathcal{O}_{\bk}^{ij} c_j$ by the post-quench state,
$\langle{\bf O}_{\bk}(t) \rangle=\langle \Psi(t)|{\bf O}_{\bk}|\Psi(t)\rangle = \sum_{\alpha,\beta,\gamma}[e^{i t \Ham_\bk}]_{\alpha\beta}\mathcal{O}_{\bk}^{\beta\gamma}[e^{- i t \Ham_\bk}]_{\gamma\alpha}$,
where $\mathcal{O}_{\bk}^{\beta\gamma}= u^\dagger_{\beta i} \mathcal{O}_{\bk}^{ij} u_{\gamma j}$.
This post-quench measurement defines an order parameter 
$\langle{\bf O}_{\bk}(t) \rangle$  on a manifold $\mathcal{M}'$.

For a finite system, the post-quench state will recur to its initial state. The momentum-time space is a $d+1$ torus
and we can consider a mapping from a point in this $d+1$ torus to a point in the order parameter space $\mathcal{M}'$.
We can classify distinct sets of the order parameter space by the $d+1$-th homotopy group $\pi_{d+1}(\mathcal{M}')$.

We demonstrate few examples where the classification of the post-quench order parameter $\pi_{d+1}(\mathcal{M}')$
had direct relation to the classification of the static Hamiltonian $\pi_{d}(\mathcal{M})$
in $d=1,2,3$ dimensions and $n$-band models with $n=2,3,4$.

\section{Two-band models}
\label{Sec:2B}
A generic two-band Hamiltonian can be written as $\mathcal{H}_{\bf k}=a_{\bf k} \mathbb{I}_{2 \times 2}+ (f_\bk, g_\bk, h_\bk) \cdot {\boldsymbol \sigma}$,
where $ {\boldsymbol \sigma}=(\sigma_x, \sigma_y,\sigma_z)$ are  the Pauli matrices.
The corresponding energy is $E_{\bf k} = a_{\bf k} \pm \sqrt{f_\bk^2+g_\bk^2+h_\bk^2}$. Since $a_{\bf k} $ just shifts the energy, the topology
of the Hamiltonian is independent of $a_{\bf k}$. For simplicity, we remove $a_{\bf k}$ in the following discussion.

In one dimensional cases, we consider a symmetry constraint restricting the Hamiltonian such that one of the Pauli matrices is forbidden.
For example, a chiral symmetry constrains the Hamiltonian  $\mathcal{S}:\to \mathcal{S}^{\dagger} \mathcal{H}_{ k}\mathcal{S}=-\mathcal{H}_{ k}$.
Then $h_{k}=0$ if $\mathcal{S}=\sigma_z$. The manifold of the Hamiltonian can be seen as a ring with the parametrization $e^{i \theta_k} =\frac{f_{k}+i g_{k} }{\sqrt{f_{k}^2+g_{k}^2}}$.
The classification for a point in $k$ to $\theta_k$ is given by the first homotopy group $\pi_1(S^1)=\mathbb{Z}$
and can be indexed by the winding number
\begin{align}
\nu &= \frac{1}{2 \pi} \int d k \frac{ d \theta_k}{ d k}  \notag\\
&= \frac{1}{2\pi}\int d k \frac{1}{f_{k}^2+g_{k}^2} [f_{k} \partial_k g_{k}-g_{k} \partial_k f_{k}].
\end{align}

In two dimensional cases, we do not consider any symmetry constraints. All the components $(f_\bk, g_\bk, h_\bk)$ are non-vanishing.
The manifold of the Hamiltonian is a two-sphere where we can parametrize it by a unit vector $\hat{d}_{\bf k} =\frac{(f_\bk, g_\bk, h_\bk)}{\sqrt{f_\bk^2+g_\bk^2+h_\bk^2}}$.
The second homotopy group classifies the Hamiltonian by $\pi_2(S^2)=\mathbb{Z}$ and can be indexed by
the Chern number
\begin{align}
C &= \frac{1}{4\pi} \int d^2 k \hat{d}_{\bf k} \cdot [\partial_{k_x}   \hat{d}_{\bf k} \times \partial_{k_y} \hat{d}_{\bf k}].
\end{align}

Now we consider an initial state $|\psi_i\rangle$ which evolves under the evolution operator $U_\bk(t)=e^{-i \mathcal{H}_{\bf k} t}$, 
$|\psi_{\bf k}(t)\rangle=U_\bk(t)   |\psi_i\rangle$. The evolution operator can be written as $U_\bk(t) = \cos (|E_{\bf k}| t) - i \mathcal{H}_{\bf k} \sin (|E_{\bf k}| t)$. 
The post-quench state will recur to its initial state at $t=2\pi/|E_{\bf k}|$.
Momentum and time  $({\bf k},t)$ form a  $d+1$-dimensional torus, where $d$ is the dimensions of the momentum space. 

\subsection{1+1 dimensions}
\begin{figure}[!htbp]
\center
\includegraphics[width=\columnwidth] {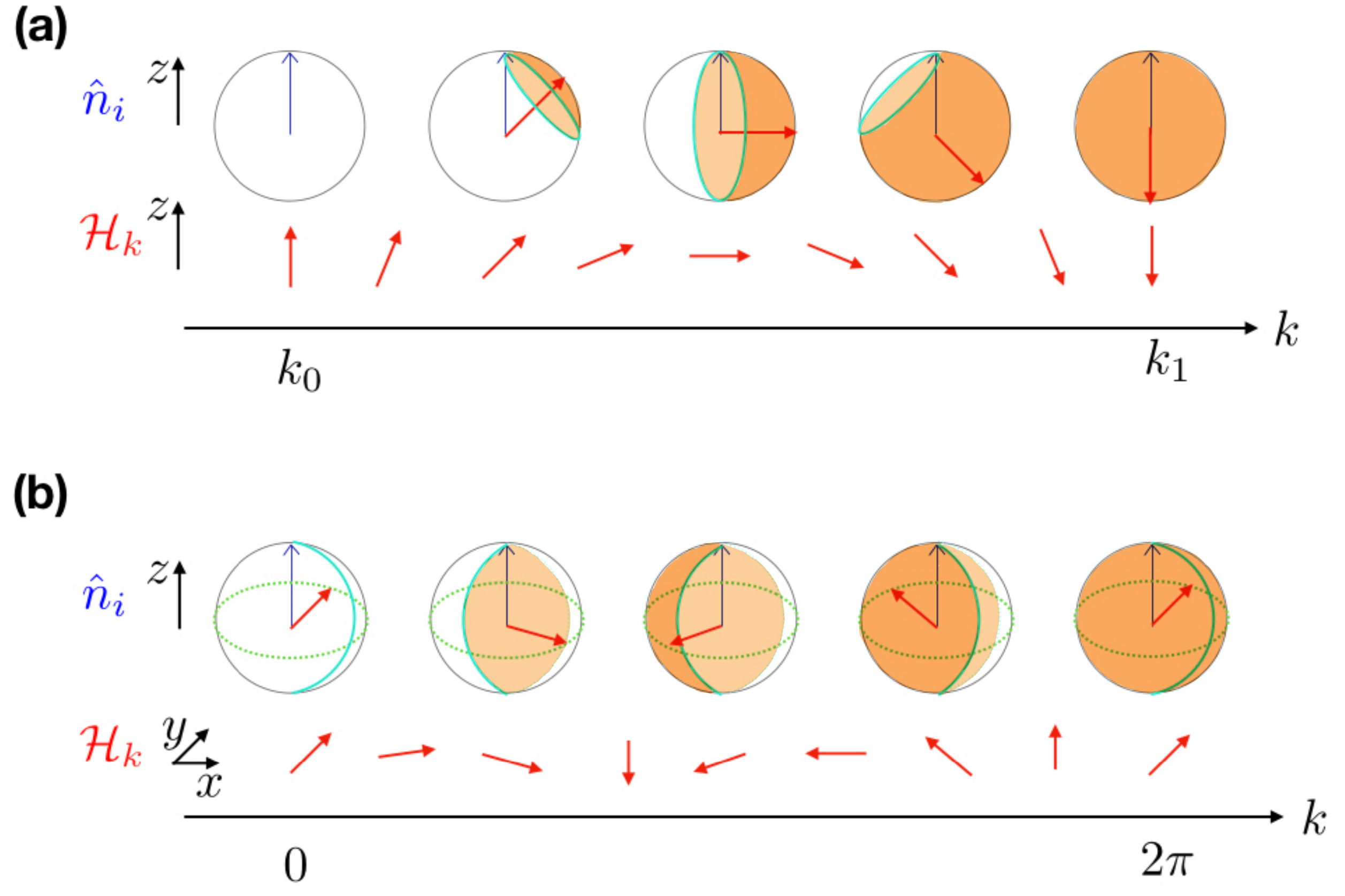}
\caption{An illustration of the topological relation between  non-vanishing winding number
of the static Hamiltonian and the non-vanishing dynamic Chern number of the post-quench pseudospin. The blue arrows corresponds to the pseudospin of the initial state.
The red arrows correspond to the directions of the pseudomagnetic field (Hamiltonian $\Ham_\bk$). The light blue (semi-)rings correspond to the trajectories of the pseudospin precession. (a)
When there are two fixed points $k_0$ and $k_1$, the trajectories of the pseudospin precession wraps the Bloch sphere from $k_0$ to $k_1$ (shaded by orange).
(b) When pseudospin of the initial state is perpendicular to the directions of the pseudomagnetic field, the trajectories of the pseudospin precession with a half period wraps the Bloch sphere from $0$ to $2\pi$ (shaded by orange).
Dashed green line shows the equator in x-y plane. }
 \label{F1}
\end{figure}

We first discuss the case when the static Hamiltonian $\Ham_k$ is in one dimension and 
the post-quench state $|\psi_k (t)\rangle $ is in 1+1 dimensions.
A pseudospin can be defined by the post-quench state as $\hat{n}_{k}(t) =\langle \psi_{k}(t)|{\boldsymbol \sigma}| \psi_{k}(t) \rangle $.
There are two possible scenarios for the post-quench pseudospin. One scenario is that there can exist fixed points such that the post-quench pseudospin is parallel or anti-parallel to the 
pseudomagnetic field (static Hamiltonian $\Ham_\bk$).
The topological invariant characterizing the post-quench pseudospin in this scenario is the 
dynamical Chern number\cite{Yang2018, Gong2017},
\begin{align}
C_{\rm dyn.} = \frac{1}{4\pi} \int^{k_{m+1}}_{k_m} dk \int_0^\pi dt \hat{n}_{ k}(t) \cdot [\partial_{k}   \hat{n}_{ k}(t) \times \partial_{t} \hat{n}_{k}(t)],
\label{eq:dc}
\end{align}
where $k_m$ and $k_{m+1}$ are two nearby fixed point in one-dimensional momentum space.
It has been shown in Refs. \onlinecite{Yang2018, Gong2017}, that the dynamic Chern number $C_{\rm dyn.} = \pm 1$ if the winding number $\nu_i$ for the Hamiltonian of the initial state 
is different than the winding number $\nu$ of the Hamiltonian $\mathcal{H}_{\bf k}$,
$\nu_i \neq \nu$.
Pictorially, one can visualize this dynamical Chern number by monitoring how many times the trajectory of the post-quench pseudospin 
wraps the Bloch sphere. We demonstrate this wrapping in Fig. \ref{F1}(a). For a given $k$, the post-quench pseudospin  precesses along the direction of the pseudomagnetic field $d_k=(f_k,g_k,0)$ with a circular
trajectory on the Bloch sphere. If the static Hamiltonian $\Ham_\bk$ has a non-trivial winding, the circular trajectory of the post-quench pseudospin can wrap the entire Bloch sphere
from $k_0$ to $k_1$, where $k_{0(1)}$ is the fixed point with the post-quench pseudospin (anti-)parallel to the pseudospin of the initial state.

Next we consider the second scenario which has not been discussed before.
This scenario has no fixed points for the post-quench pseudospin. 
In general, the  circular trajectories on the Bloch sphere of the post-quench pseudospin from $k=0$ to $k=2\pi$  do
not wrap the entire Bloch sphere. Here $k\in [0, 2\pi]$ is the Brillouin zone (BZ). 
However, when the  pseudospin of the initial state is perpendicular to the direction of the pseudomagnetic field, 
the circular trajectories on the Bloch sphere of the post-quench pseudospin from $k=0$ to $k=2\pi$ can wrap the entire Bloch sphere [see Fig \ref{F1}(b)].

Without loss of generality, we consider the static Hamiltonian $\Ham_\bk = \hat{f}_\bk \sigma_x + \hat{g}_\bk \sigma_y$ and the initial state $|\psi_i\rangle =(1,0)^{\rm T}$.
Here we normalize the static Hamiltonian $\hat{f}_\bk^2 + \hat{g}_\bk^2=1$.
The post-quench state is $|\psi_\bk(t)\rangle =(\cos t,- i (\hat{f}_\bk +i \hat{g} _\bk) \sin t)^{\rm T}$
and the post-quench pseudospin is $\hat{n}_\bk(t)=(\hat{g}_\bk \sin 2 t,\hat{f}_\bk \sin 2t, \cos 2t)$.
Since there is no fixed point, we need to integrate out the entire momentum space for the corresponding dynamical Chern number, which is defined as
\begin{align}
C_{\rm dyn.}' &= \frac{1}{4\pi} \int^{2 \pi}_{0} dk \int_0^{\pi/2} dt \hat{n}_{\bf k}(t) \cdot [\partial_{k}   \hat{n}_{\bf k}(t) \times \partial_{t} \hat{n}_{\bf k}(t)] \notag\\
&=\frac{-1}{2 \pi} \int^{\pi/2}_0\sin 2t \int_0^{2 \pi} dk  [f_k \partial_k g_k -g_k \partial_k f_k] \notag\\
&=-\nu.
\label{eq:dc2}
\end{align}
Because of the intrinsic symmetry, $\hat{n}_{\bf k}(t) \cdot [\partial_{k}   \hat{n}_{\bf k}(t) \times \partial_{t} \hat{n}_{\bf k}(t)]
=\hat{n}_{\bf k}(-t) \cdot [\partial_{k}   \hat{n}_{\bf k}(-t) \times \partial_{t} \hat{n}_{\bf k}(-t)]$,
we integrate half of the period ($0$ to $\pi/2$) of the post-quench pseudospin to have the non-vanishing dynamical Chern number.
Fig. \ref{F1}(b) shows how the trajectories of the pseudospin precess with half period wraps the Bloch sphere when there is nonvanishing winding number of the static Hamiltonian
$\Ham_\bk$.

\begin{figure}[!htbp]
\center
\includegraphics[width=\columnwidth] {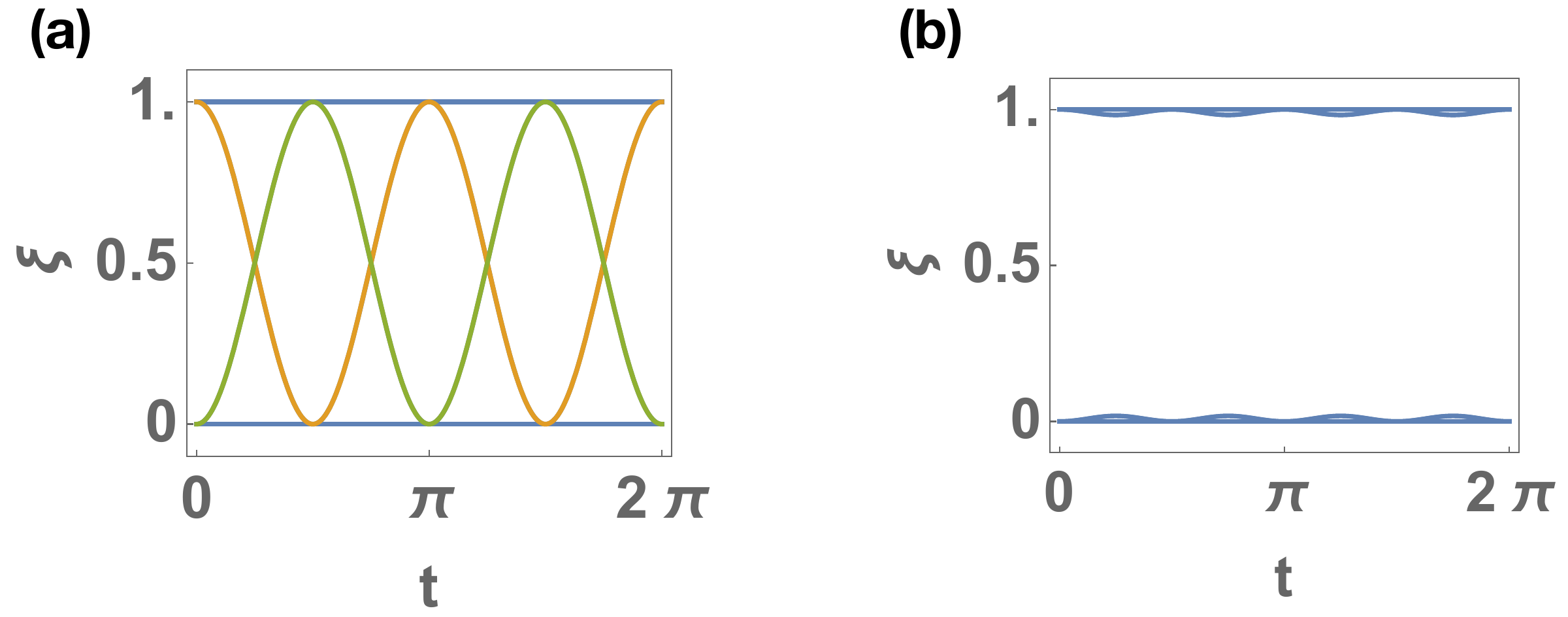}
\caption{Entanglement spectrum $\xi(t)$ for a real space bipartition with (a) $t_1/t_2=0.5$. The green (orange) line indicates the corresponding eigenstate localized 
at the left (right) entangling boundary between subsystems $A$ and $B$. (b) $t_1/t_2= 2$. There are no crossings in the entanglement spectrum. }
 \label{Fig:EE}
\end{figure}

To demonstrate the topological property of the post-quench states, we consider the Su-Schrieffer-Heeger model,
where $f_k = t_1+t_2 \cos k$ and $g_k = t_2 \sin k$. When $|t_1/t_2|<1$ the winding number $\nu=1$ and $|t_1/t_2|>1$ the winding number $\nu=0$. 
One way to characterize topological property of the post-quench states is the entanglement spectrum.
One can consider a spatial bipartition (A and B subsystems) and construct the reduced density matrix $\rho_A(t) ={\rm Tr}_B |\Psi(t) \rangle \langle \Psi(t)|
=\mathcal{N}^{-1} e^{-H_A(t)}$, where $H_A(t)$ is the entanglement Hamiltonian and $\mathcal{N}$ is the normalization condition such that ${\rm Tr}_A\rho_A(t)=1 $.
In free-fermion systems, the spectrum of the entanglement Hamiltonian can be directly extracted from the correlation matrix $C^A_{x,x'}(t)= \langle \Psi (t)| c^\dagger_x c_{x'} | \Psi (t) \rangle$,
with $x,x'$ in the subsystem $A$ \cite{Peschel2003, PYC2014}.
The entanglement spectrum $\xi(t)$ is the defined as the eigenvalue of the reduced density matrix $C^A_{x,x'}(t)$.
In the first scenario where there are fixed points, it is shown in Ref. \onlinecite{Gong2017} that the entanglement spectrum has crossings if the the dynamic Chern number is non-vanishing.
Here, we demonstrate that in the second scenario, the entanglement spectrum also has crossings if the dynamic Chern number is non-vanishing.
Fig. \ref{Fig:EE}(a) shows when $t_1/t_2=0.5$, the mid-gap states in the entanglement spectrums crosses.
These mid-gap states are localized at the entanglement boundary. This is the bulk boundary correspondence in the entanglement Hamiltonian. If 
the dynamic Chern number is non-vanishing, there are robust boundary modes in the entanglement Hamiltonian. 
On the other hand, if the dynamic Chern number is zero, there is no localized mid-gap states as shown in Fig. \ref{Fig:EE}(b). 

In two-band models, the post-quench pseudospin $\hat{n}_\bk (t)$ contains same 
information as the post-quench projector $P_k(t) = |\psi_k \rangle \langle \psi_k| = \frac{1}{2} (1+ \tilde{n}_k (t)\cdot \boldsymbol{\sigma})$,
where $\tilde{n}_k (t) = (\hat{g}_k \sin 2 t,-\hat{f}_k \sin 2t, \cos 2t)$.
Hence the topological invariant computed from the post-quench pseudospin
can reflect the topology of the post-quench state and can reveal its property in the entanglement spectrum.
In $1+1$ dimensional cases, the Chern number $C$ computed from the post-quench state
is $C=-C'_{\rm dym.}=\nu$. 
This indicates that the number of crossings of the mid-gap states in the entanglement Hamiltonian 
is related to the dynamical Chern number of the post-quench order parameter.
Since the dynamical Chern number is computed from $t=0$ to $t=\pi/2$,
the number of the crossings in the entanglement spectrum in the region $t\in[0,\pi/2]$ equals to the dynamical Chern number
 due to the bulk boundary correspondence.


\subsection{2+1 dimensions}

For the case that the static Hamiltonian is in two dimensions and the post-quench state is in 2+1 dimensions, 
the post-quench pseudospin is defined as $\hat{n}_{\bf k}(t) =\langle \psi_{\bf k}(t)|{\boldsymbol \sigma}| \psi_{\bf k}(t) \rangle $ on the Bloch sphere.
One can consider a mapping from $(t, k_x, k_y)$ to $\hat{n}_k$ that can be classified by the third homotopy group $\pi_3(S^2)$.
The topological index is characterized by the Hopf number\cite{Wilczek1983},
\begin{align}
\chi = \int d^2 k dt {\bf F_\bk (t)} \cdot {\bf A_\bk (t)},
\end{align}
where $F^i_\bk (t) = \frac{1}{8\pi} \epsilon^{ijk} \hat{n}_{\bf k}(t) \cdot [\partial_{j}   \hat{n}_{\bf k}(t) \times \partial_{k} \hat{n}_{\bf k}(t)]$
 and $A^i_\bk(t)$ is the Berry connection satisfying $F^i_\bk (t)= \epsilon^{ijk} \partial_{j} A^k_\bk (t)$.
In a two-band model, the Berry connection $A^i_\bk(t)$ and Berry flux $ F^i_\bk (t)$ can be computed from the post-quench state
$A^i_\bk(t)= i \langle \psi_\bk(t) | \partial_i\psi_\bk(t) \rangle $ and
$ F^i_\bk (t) = \frac{1}{2\pi}\epsilon^{ijk} \langle \partial_j \psi_\bk(t)| \partial_k \psi_\bk(t)\rangle $. 
It has been shown in Refs. \onlinecite{Wang2017,Tarnowski2017}, that the Hopf number is non-vanishing if the Chern number of the static Hamiltonian is non-zero.

\begin{figure}[!htbp]
\center
\includegraphics[width=\columnwidth] {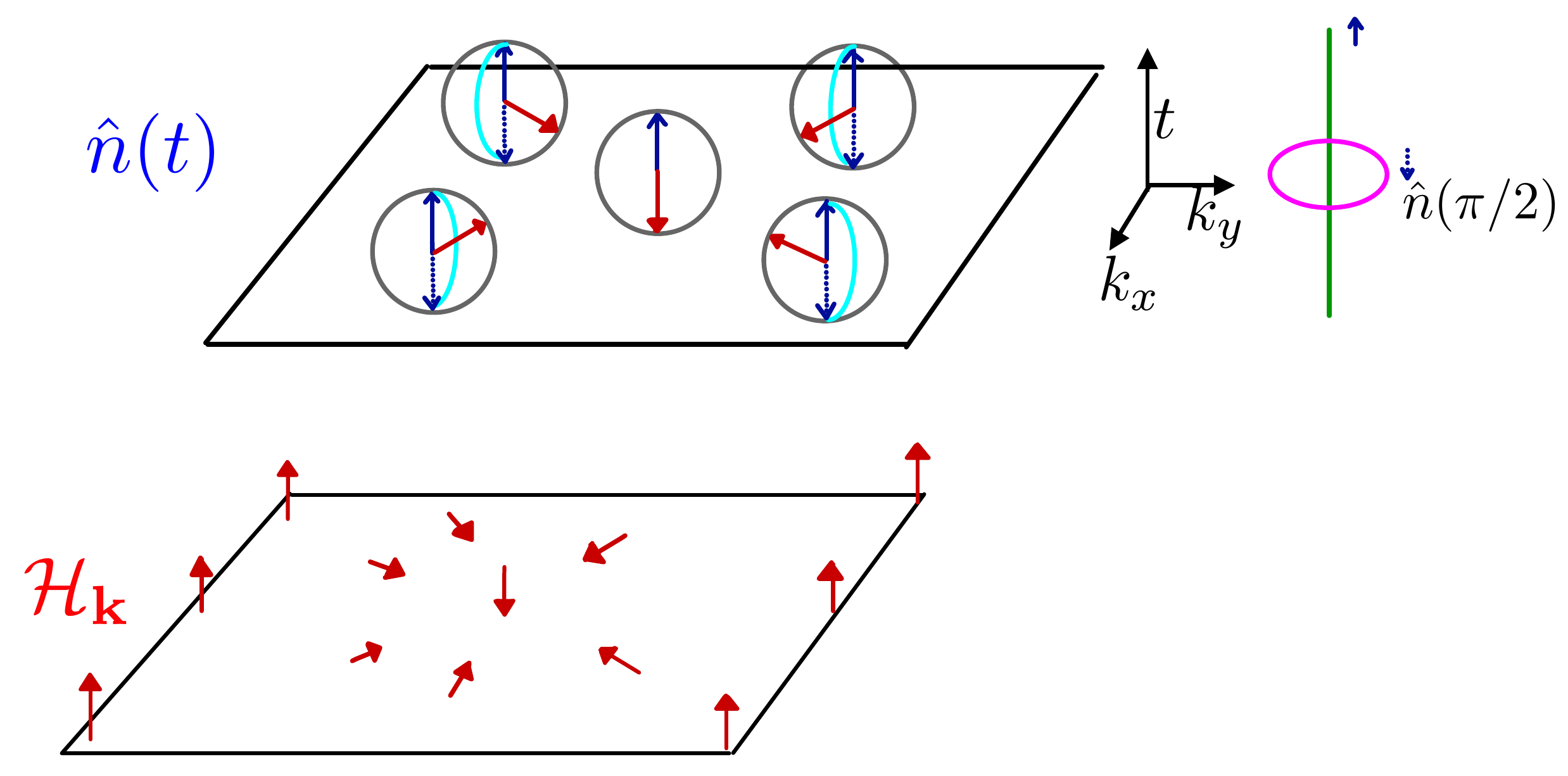}
\caption{An illustration of the topological relation between non-vanishing Chern number
of the static Hamiltonian and the non-vanishing Hopf number of the post-quench pseudospin. 
The blue arrows corresponds to the pseudospin $\hat{n}_\bk(t)$ of the initial state [upper panel].
The red arrows correspond to the direction of the pseudomagnetic field $d_\bk=(f_\bk,g_\bk,h_\bk)$ [lower panel]. 
The light blue arcs correspond to the trajectories of the pseudospin $\hat{n}_\bk(t)$ precession from $t=0$ to $\pi/2$ [upper panel].
The dashed blue arrows are the pseudospin at $t=\pi/2$ [upper panel].
In the upper right panel, the green line corresponds to the inverse mapping from the pseudospin pointing to the north pole at center
to the momentum-time space. The pink ring corresponds to the inverse mapping from the pseudospin pointing to the south pole at $t=\pi/2$ 
to the momentum-time space. The green line and the pink ring form a Hopf link that relates to non-vanishing Hopf number of the post-quench pseudospin.}
 \label{F2}
\end{figure}

The relation between the non-vanishing Chern number of the static Hamiltonian and
the non-vanishing Hopf number is illustrated in Fig. \ref{F2}.
If the static Hamiltonian has non-vanishing Chern number, the direction of pseudomagnetic field $d_\bk = (f_\bk, g_\bk, h_\bk)$
forms a skyrmion texture. The post-quench pseudospin precesses under the pseudomagnetic field. 
For a given direction of the pseudospin, there is an inverse mapping from $\hat{n}_\bk(t) \to (k_x, k_y, t)$ such that the trajectory in 
the momentum-time space is a close loop. If the post-quench pseudospin has non-vanishing Hopf number,
the inverse mapping of two different pseudospins form a Hopf link in the momentum-time space\cite{Wang2017,Tarnowski2017}. 
As demonstrate in  Fig. \ref{F2}, the pseudospin at center is anti-parallel to the pseudomagnetic field and does not precess.
The inverse mapping is a line along the $t$-axis. For the pseudospins perpendicular to the pseudomagnetic fields which
are pointing to the center, these pseudospins will precess to the south pole at $t=\pi/2$. The inverse mapping of 
these pseudospins pointing to the south pole is a ring at $t=\pi/2$ plane that encircling the inverse mapping of the pseudospins pointing along the north pole.
These two trajectories in the momentum-time space form a Hopf link that relates to a non-vanishing Hopf number of the post-quench pseudospin.

\begin{figure}[!htbp]
\center
\includegraphics[width=\columnwidth] {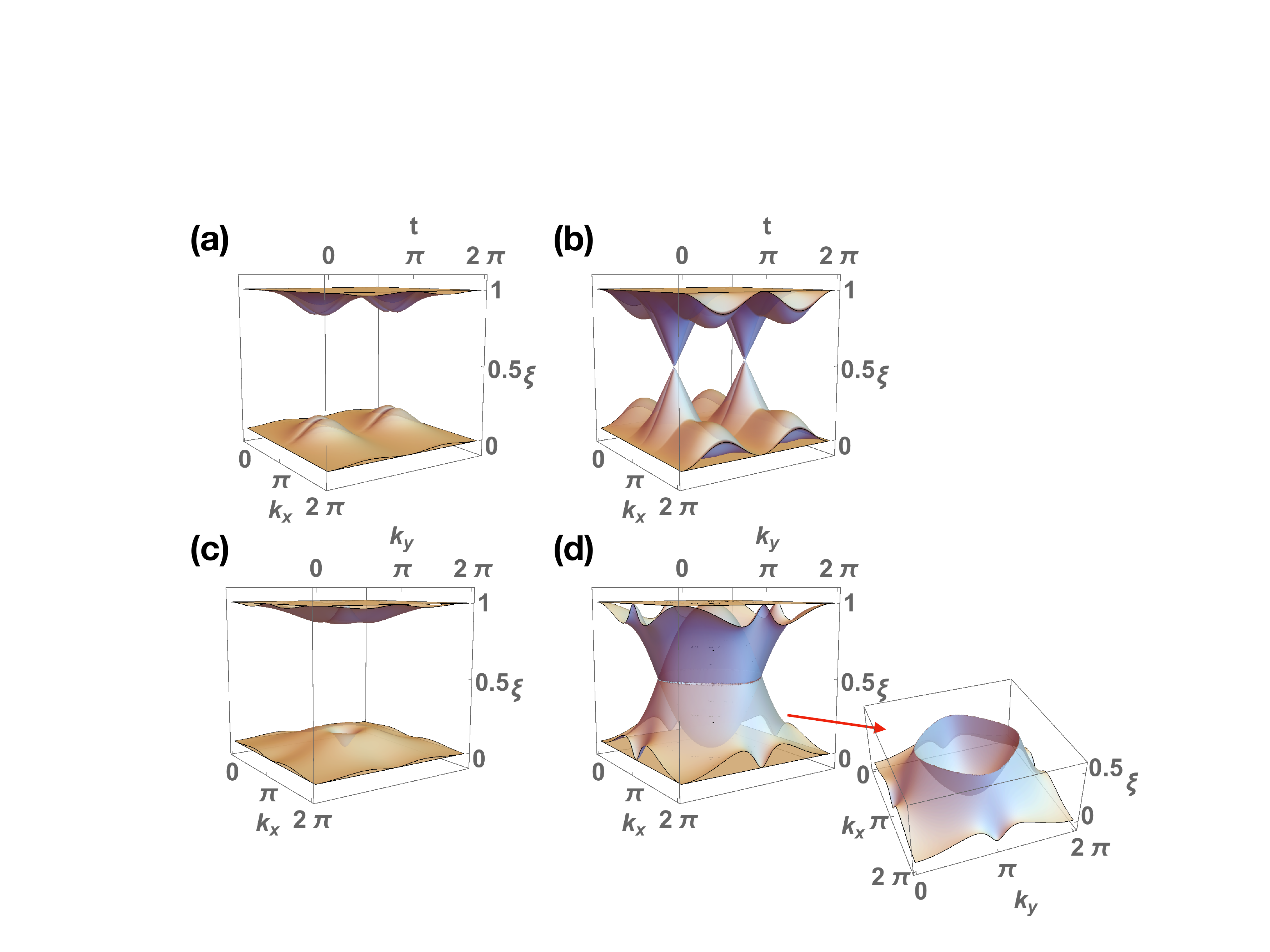}
\caption{Entanglement spectrum $\xi(k_x,t)$ for a real space bipartition with (a) $(t_1, M)=(1,2.5)$, (b) $(t_1, M)=(1,0.5)$.
Entanglement spectrum $\xi(k_x,k_y)$ for a frequency space bipartition with (c) $(t_1, M)=(1,2.5)$, (d) $(t_1, M)=(1,0.5)$.
The right panel in (d) shows the mid-gap states from a ring.}
 \label{Fig:EES2}
\end{figure}

Now we will show the relation of the entanglement spectrum of the post-quench state
and its corresponding Hopf number characterizing the post-quench pseudospin. To be specific, we consider the Hamiltonian with 
$f_\bk=t_1 \sin k_x$,
$g_\bk={t_1 \sin k_y}$, and
$h_\bk={M}+\cos k_x + \cos k_y$.
The Chern number  of this static Hamiltonian is $|C| =  1$ when $0< |M/t_1|< 2 $ and $C=0$ otherwise.
We consider the entanglement spectrum of the post-quench state evolved from $| \psi_i \rangle =(1,0)^{\rm T}$.
For simplicity, we flatten the Hamiltonian $|E_k|=1$ such that the period of the post-quench state is $2 \pi$. 
For a real space bipartition, the entanglement spectrum has no crossings if the Hopf number is zero [Fig. \ref{Fig:EES2}(a)]
and has two cones when the Hopf number $|\chi|=1$  [Fig. \ref{Fig:EES2}(b)].
On the other hand, we can also consider a frequency space bipartition.
The post-quench state in the frequency space is $\psi_\bk(\omega) = \int_0^{2\pi} dt e^{i \omega t} \psi_\bk(t)$,
with $\omega \in [0,1]$. The frequency space bipartition we considered is that $A:  \omega \in [0,0.5]$ and  $B: \omega \in [0.5,1]$.
The entanglement spectrum has no crossing if the Hopf number is zero [Fig. \ref{Fig:EES2}(c)].
For the case Hopf number $|\chi|=1$, the mid-gap states in the entanglement spectrum form a ring [Fig. \ref{Fig:EES2}(d)].
This ring in the entanglement spectrum is similar as the boundary Fermi ring in the Hopf insulators \cite{Liu2017,Moore2008, Deng2013}.

Up to date, the relation between the Hopf number and
 the number of boundary Fermi rings in the Hopf insulators has not been established in the literature.
 It has been roughly discussed in Ref. \onlinecite{Deng2013} that 
there are more surfaces states when the absolute value of the Hopf number becomes larger.
One should be noticed that in the Hopf insulators, all the Chern numbers computed from the three
two dimensional tori embeded in $T^3$ are zero. This indicates that there is no chiral modes on three boundaries
due to vanishing of the Chern numbers in three directions.
However, non-vanishing Hopf number generates gapless boundary states from the bulk boundary correspondence.
Heuristically, the Fermi ring can be understood from a skyrmion texture of the pseudospin $\hat{n}(t)$ in the $(k_r, t)$ plane,
where $k_r = \sqrt{k_x^2+k_y^2}$. In Fig. \ref{F2}, at $k_r=0$ and $k_r = k_{\rm boundary}$, the pseudospin always points to the north pole.
The trajectory in the momentum-time space where the pseudospin points to the south pole forms a ring [pink ring in Fig. \ref{F2}].
Hence for a fixed $\theta_k = \tan^{-1} k_x/k_y$, the pseudospin $\hat{n}(t)$ has a space-time skyrmion texture in the $(k_r, t)$ plane
and leads to one chiral boundary mode in entanglement Hamiltonian for a frequency space bipartition.
The Fermi ring in the entanglement Hamiltonian is the collection of the chiral boundary modes from $\theta_k=0$ to $2\pi$.

\section{Three-band models: spin-1 models}\label{Sec:S1}

A generic three-band model can be written as 
$\Ham_{\bk} = a_{\bk} \mathbb{I}_{3 \times 3} + {\bf b}_{\bk} \cdot {\boldsymbol\lambda}$,
where $\lambda_i, i=1,\cdots 8$ are Gell-Mann matrices spanning the Lie algebra of the SU(3) in the defining representation. 
We remove $a_{\bk}$ in the Hamiltonian since it just shifts the energy level.
One can flatten the Hamiltonian by using the eigenstate projectors in terms of Gell-Mann matrices as\cite{Barnett2012,Lee2015}
\begin{align}
P_{\bk, \alpha}= | \psi_{\bk, \alpha}\rangle \langle \psi_{\bk, \alpha}| = \frac{1}{3}  (1 + \sqrt{3} {\bf n}_{\bk, \alpha} \cdot  {\boldsymbol\lambda}),
\end{align}
where two conditions ${\rm Tr} P_{\bk, \alpha}=1$ and $P_{\bk, \alpha}^2=P_{\bk, \alpha}$ constrain
the ${\bf n}_{\bk, \alpha} $ vectors to be a unit vector on $S^7$.
The $ {\bf n}_{\bk} $ vector describes the manifold of the static Hamiltonian with higher dimension than $S^1$  and $S^2$.
The homotopy group is zero in one and two dimensions, $\pi_1(S^7)=0$ and  $\pi_2(S^7)=0$. 
To have the non-trivial homotopy group, we need to constrain the Hamiltonian to have the spin-1 structure,
$\Ham_{\bk} = {\bf d}_{\bk} \cdot {\bf S}$, where $ {\bf d}_{\bk} = (f_{\bk}, g_{\bk}, h_{\bk})$
and $ {\bf S}=(S_x, S_y, S_z)$ are chosen from the linear combination of the Gell-Mann which satisfy the SU(2) sub-algebra.
In the following discussion, we consider the representation of $ {\bf S}$ to be
\begin{align}
S_x &= \frac{1}{\sqrt{2}}\left(\begin{array}{ccc}0 & 1 & 0 \\1 & 0 & 1 \\0 & 1 & 0\end{array}\right), \quad
S_y = \frac{1}{\sqrt{2}}\left(\begin{array}{ccc}0 & -i & 0 \\i & 0 & -i \\0 & i & 0\end{array}\right), \notag\\
S_z &= \left(\begin{array}{ccc}1 & 0 & 0 \\ 0 & 0 & 0 \\0 & 0 & -1\end{array}\right).
\end{align}
The corresponding energies are $E_{0}= 0$ and $E_{\bk,\pm}= \pm \sqrt{f_{\bk}^2+g_{\bk}^2 +h_{\bk}^2}$. 

The spin-1 models have the same classifications of static Hamiltonian as two-band models.
In one dimensional cases,
if we eliminate one of the $S_i$ (e.g., we set $h_{\bk}=0$) in the Hamiltonian, the 
Hamiltonian can be classified by the first homotopy group $\pi_1(S^1)$ with the winding number
$\nu = \frac{1}{2\pi}\int d k  [\hat{f}_{k} \partial_k \hat{g}_{k}-\hat{g}_{k} \partial_k \hat{f}_{k}]$,
where $\hat{f}_{k}=\frac{f_{k}}{\sqrt{f_{k}^2+g_{k}^2}}$ and $\hat{g}_{k}=\frac{g_{k}}{\sqrt{f_{k}^2+g_{k}^2}}$.
In two dimensional cases, we assume all the components in ${\bf d}_{\bk}$ are non-vanishing.
The Hamiltonian can be classified by the second homotopy group $\pi_2(S^2)$  with the Chern number
$C= \frac{1}{4\pi}\int d^2 k \hat{d}_{\bf k} \cdot [\partial_{k_x}   \hat{d}_{\bf k} \times \partial_{k_y} \hat{d}_{\bf k}]$,
where 
$\hat{d}_{\bf k} =\frac{{\bf d}_{\bf k}}{|{\bf d}_{\bf k}|}$.

The evolution operator of the spin-1 models satisfies the Rodrigues rotation formula\cite{Curtright2014},
\begin{align}
U_\bk(t)&=\exp [- i {\Ham}_{\bk} t] \notag\\
&= \mathbb{I}_{3 \times 3} - i {\Ham}_{\bk} \sin (|E_{\bk,\pm}|t)+  \Ham_{\bk}^2 (\cos (|E_{\bk,\pm}|t) -1).
\label{Eq:S1}
\end{align} 
We have $U_\bk(2\pi/ |E_{\bk,\pm}|) =U_\bk(0)$. The post-quench state will recur to the initial state at $t=2\pi/ |E_{\bk,\pm}|$.


Here we define the post-quench pseudospin $\hat{s}_\bk = \langle \psi_\bk(t) | {\bf S} |  \psi_\bk(t) \rangle$.
Unlike two-band models, the post-quench pseudospin in the spin-1 models is not guarantee to be a unit vector.
For example, $\hat{s}= \langle 010 | {\bf S} | 010\rangle=(0,0,0)$ is a null vector, where $(010)$ is a shorthand notation for a three-dimensional state $|\psi \rangle =(0,1,0)^{\rm T}$.
We need to restrict the initial state such that the post-quench pseudospin is a unit vector.
In the representation of the ${\bf S}$ we chosen, the initial state can be either $|\psi_i\rangle=(1,0,0)^{\rm T}$ or $(0,0,1)^{\rm T}$ to maintain
the norm of the post-quench pseudospin to be one.

To summarize, in order to have non-trivial topology of both the static Hamiltonian and post-quench order parameter in a three-band model, 
we focus on spin-1 models with a given initial state either $|\psi_i\rangle=(1,0,0)^{\rm T}$ or $(0,0,1)^{\rm T}$.


\subsection{1+1 dimensions}

In 1+1 dimensional cases, we consider the static Hamiltonian 
$\Ham_k= f_k S_x + g_k S_y $ such that the topology of the static Hamiltonian
can be indexed by the winder number $\nu$. 
Let us assume the initial state is $| \psi_i \rangle = (1,0,0)^{\rm T}$. To further simplify the calculation and have a more compact form,
we normalize the Hamiltonian $\hat{f}_k^2+\hat{g}_k^2=1$.
The post-quench state evolved under the evolution operator from Eq. (\ref{Eq:S1}) is
\begin{align}
|\psi_k (t) \rangle =U_k(t) | \psi_i \rangle=  \left(\begin{array}{c}\cos ^2 \frac{t}{2}(\hat{f}_k- i \hat{g}_k) \\ - i \sin t/\sqrt{2} \\-\sin^2 \frac{t}{2}(\hat{f}_k+i \hat{g}_k)\end{array}\right).
\label{Eq:11}
\end{align}

The post-quench pseudospin is $\hat{s}(k,t) =( g_k \sin t ,- f_k \sin t , \cos t)$ pointing on a two-sphere $S^2$.
We can now define the dynamical Chern number characterizing the topology of the post-quench pseudospin as
\begin{align}
C_{\rm dyn.}'&=\frac{1}{4 \pi} \int_0^{2\pi} dk \int_{0}^{\pi} dt \hat{s} \cdot (\partial_k \hat{s} \times \partial_t \hat{s}) \notag\\
&=\frac{1}{4 \pi} \int^\pi_0 dt \sin t \int_0^{2\pi} dk [f_k \partial_k g_k - g_k\partial_k f_k]  \notag\\
&=\nu.
\end{align}
Similar as the two-band cases, there is an intrinsic symmetry such that
$\hat{n}_{\bf k}(t) \cdot [\partial_{k}   \hat{n}_{\bf k}(t) \times \partial_{t} \hat{n}_{\bf k}(t)]
=\hat{n}_{\bf k}(-t) \cdot [\partial_{k}   \hat{n}_{\bf k}(-t) \times \partial_{t} \hat{n}_{\bf k}(-t)]$.
We integrate half of the period ($0$ to $\pi$) 
to have the non-vanishing dynamical Chern number.



\begin{figure}[!htbp]
\center
\includegraphics[width=\columnwidth] {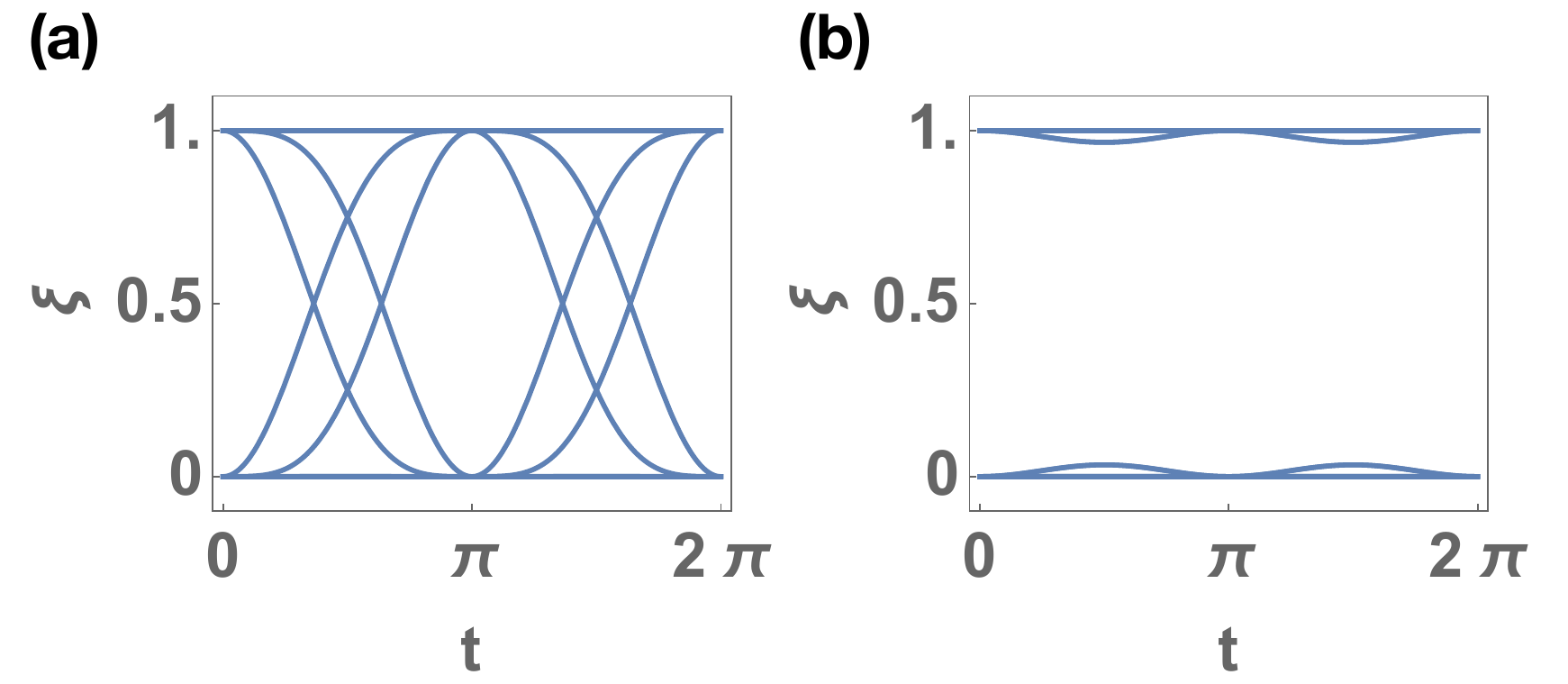}
\caption{Entanglement spectrum $\xi(t)$ computed from the post-quench state defined in Eq. (\ref{Eq:11}) for a real space bipartition with (a) $(t_1, t_2)=(0.5,1)$, (b) $(t_1, t_2)=(1,0.5)$.}
 \label{ee1}
\end{figure}

In the case $\hat{f}_k=\frac{ t_1+t_2 \cos k}{\sqrt{( t_1+t_2 \cos k)^2+t_2^2 \sin^2 k}}$ and 
$\hat{g}_k=\frac{t_2 \sin k}{\sqrt{( t_1+t_2 \cos k)^2+t_2^2 \sin^2 k}}$, we have $C_{\rm dym.}'=\nu=1$ when $|t_1/t_2|<1$
and $C_{\rm dym.}'=\nu=0$ when $|t_1/t_2|>1$.
We can also computed the Chern number of the post-quench state [Eq. (\ref{Eq:11})]directly
\begin{align}
C&= \frac{1}{2\pi i} \int_0^{2\pi} dk \int_{0}^{\pi} dt \langle \partial_t \psi_k(t)|  \partial_k \psi_k(t) \rangle-  \langle \partial_k \psi_k(t)|  \partial_t \psi_k(t) \rangle \notag\\
&=\frac{1}{2 \pi} \int^\pi_0 dt \sin t \int_0^{2\pi} dk [f_k \partial_k g_k - g_k\partial_k f_k]  \notag\\
&=2 C'_{\rm dyn.}.
\end{align}
The number of crossings of the mid-gap states in the entanglement Hamiltonian is direct 
related to the (dynamical) Chern number. Since the (dynamical) Chern number is computed from $t=0$ to $t=\pi$,
the number of the crossings in the entanglement spectrum in the region $t\in[0,\pi]$ equals to the Chern number
 due to the bulk boundary correspondence.
The entanglement spectrum of the post-quench state is shown in Fig. \ref{ee1}.
When the Chern number $C=2 C'_{\rm dyn.}=2$, the entanglement spectrum has two crossings from $t=0$ to $t=\pi$.
On the other hand, when the Chern number is vanishing, the entanglement spectrum does not  have crossings.

Since the crossings in the entanglement spectrum is directly related to the topology of the post-quench state 
instead of  the post-quench order parameter, it is interesting to check the Chern number of
the post-quench state with vanishing post-quench order parameter $\hat{s}=0$.
We consider the initial state $|\psi_i \rangle=  (0,1,0)^{\rm T}$ with the inital order parameter $\hat{s}=0$.
The post-quench state has the form
\begin{align}
\psi_k(t) = U_k(t) (0,1,0)^{\rm T} =  \left(\begin{array}{c} - \frac{i}{\sqrt{2}} \sin t (\hat{f}_k-i \hat{g}_k)  \\ \cos t \\ - \frac{i}{\sqrt{2}} \sin t (\hat{f}_k+i \hat{g}_k) \end{array}\right).
\end{align}
The post-quench order parameter  remains a null vector and 
the Chern number of the post-quench state is zero.

\subsection{2+1 dimensions}
In 2+1 dimensional cases, the normalized Hamiltonian can be described as $\Ham_\bk = \hat{d}_\bk \cdot {\bf S}$ where
the unit vector $\hat{d}_\bk=(\hat{f}_\bk,\hat{g}_\bk,\hat{h}_\bk)$ characterizes the topology of the static Hamiltonian.
The post-quench state is
\begin{align}
&|\psi (k_1,k_2, t) \rangle \notag\\
&= \left(\begin{array}{c}1+(\cos t -1 )(1-\frac{1}{2} (f_\bk^2 +g_\bk^2))-i h_\bk \sin t  \\\frac{1}{\sqrt{2}} (f_\bk+ i g_\bk) (h_\bk (\cos t - 1) - i \sin t) \\ \frac{1}{2} (\cos t -1 )(f_\bk+i g_\bk)^2\end{array}\right).
\label{Eq:13}
\end{align}
The post-quench pseudospin is 
$\hat{s}_\bk(t) = (f_\bk h_\bk(1-\cos t)+g_\bk \sin t, g_\bk h_\bk(1-\cos t)- f_\bk \sin t, \cos t + h_\bk^2 (1-\cos t))$ on $S^2$. 

The topology of this post-quench pseudospin is characterized by a Hopf fibration $S^3 \to S^2$.
To compute the Hopf number, we first consider a mapping of the post-quench pseudospin from $S^2$ to $S^3$.
Then the combined mapping from the momentum-time space $T^3$ to $S^3$ is characterized by $\pi_3(S^3)$ and can be indexed by the three-dimensional winding number. 
I.e., the Hopf number can be computed from the three-dimensional winding number.
The mapping is $\hat{s}_\bk \to (\eta_{\uparrow\bk}, \eta_{\downarrow\bk})$ with $\eta_{\uparrow\bk}$ and $\eta_{\downarrow\bk}$ being two complex numbers 
 satisfying $ |\eta_{\uparrow\bk}|^{2}+|\eta_{\downarrow\bk}|^{2}=1$.
\begin{align}
s_{\bk x} + i s_{\bk y} = 2 \eta_{\uparrow\bk} \bar{\eta}_{\downarrow\bk}, \quad s_{\bk z} = |\eta_{\uparrow\bk}|^{2}-|\eta_{\downarrow\bk}|^{2}.
\end{align}
This maps $S^2$ to $S^3$. Defining $\eta_{\uparrow\bk}=n_{1}+i n_{2}$ and $\eta_{\downarrow\bk}=n_{3}+i n_{4}$,
we have $(n_{1}, n_{2}, n_{3}, n_{4})=(\cos \frac{t}{2}, h_\bk \sin \frac{t}{2},g_\bk  \sin \frac{t}{2},f_\bk  \sin \frac{t}{2})$.

\begin{figure}[!htbp]
\center
\includegraphics[width=7 cm] {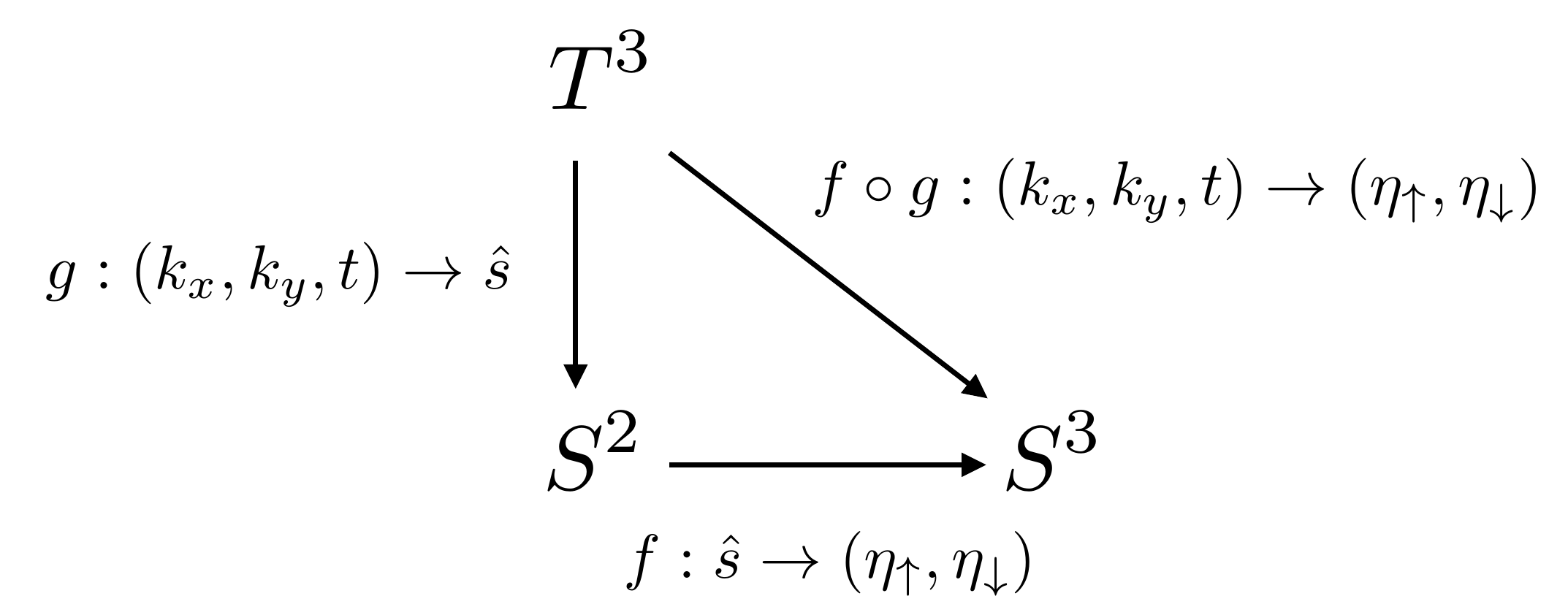}
\caption{A combined mapping from the momentum-time torus to $(\eta_\uparrow, \eta_\downarrow)$. This mapping 
can be characterized by the three-dimensional winding number in Eq. (\ref{Eq:3nu}).}
 \label{Fig:map}
\end{figure} 

The Hopf number can be computed from the mapping $T^3 \to S^3 \to S^2$ [see Fig. \ref{Fig:map}] by
\begin{align}
\chi &= \frac{1}{2 \pi^2} \int_0^{2\pi} dt  \int_{\rm BZ} d^2k \epsilon_{\mu\nu\rho\tau} n_\mu \partial_{k_x} n_\nu \partial_{k_y} n_\rho  \partial_{t} n_\tau \notag\\
 &=\frac{1}{4\pi^2} \int_0^{2\pi} dt \sin^2 \frac{t}{2} \int_{\rm BZ} d^2k \hat{d} \cdot (\partial_{k_x} \hat{d} \times \partial_{k_y} \hat{d}) \notag\\
 &=\frac{1}{4 \pi}  \int_{\rm BZ} d^2k \hat{d}_\bk \cdot (\partial_{k_x} \hat{d}_\bk \times \partial_{k_y} \hat{d}_\bk) \notag\\
 &=C.
 \label{Eq:3nu}
\end{align}
Here $C=\frac{1}{4 \pi}  \int d^2k \hat{d} \cdot (\partial_{k_x} \hat{d} \times \partial_{k_y} \hat{d})$
is the Chern number characterizing the static Hamiltonian. 
Hence we demonstrate the relation between the 
Hopf number of the post-quench pseudospin and the Chern number of the static Hamiltonian $\Ham_\bk=\hat{d}_\bk \cdot {\bf S}$ by $\chi = C$.


The static Hamiltonian of the spin-$1$ models with non-vanishing Chern number has been discussed in Ref. \onlinecite{Go2013}.
Here we consider the static Hamiltonian with 
\begin{align}
&\hat{f}_\bk=\frac{t_1 \sin k_x}{\sqrt{t_1^2 (\sin^2 k_x + \sin^2 k_y)+(M+\cos k_x + \cos k_y)^2}}, \notag\\
&\hat{g}_\bk=\frac{t_1 \sin k_y}{\sqrt{t_1^2 (\sin^2 k_x + \sin^2 k_y)+(M+\cos k_x + \cos k_y)^2}}, \notag\\
&\hat{h}_\bk=\frac{M+\cos k_x + \cos k_y}{\sqrt{t_1^2 (\sin^2 k_x + \sin^2 k_y)+(M+\cos k_x + \cos k_y)^2}}.
\end{align}
The Hopf number  $| \chi| =  1$ when $0< |M/t_1|< 2 $ and $\chi=0$ otherwise.

\begin{figure}[!htbp]
\center
\includegraphics[width=\columnwidth] {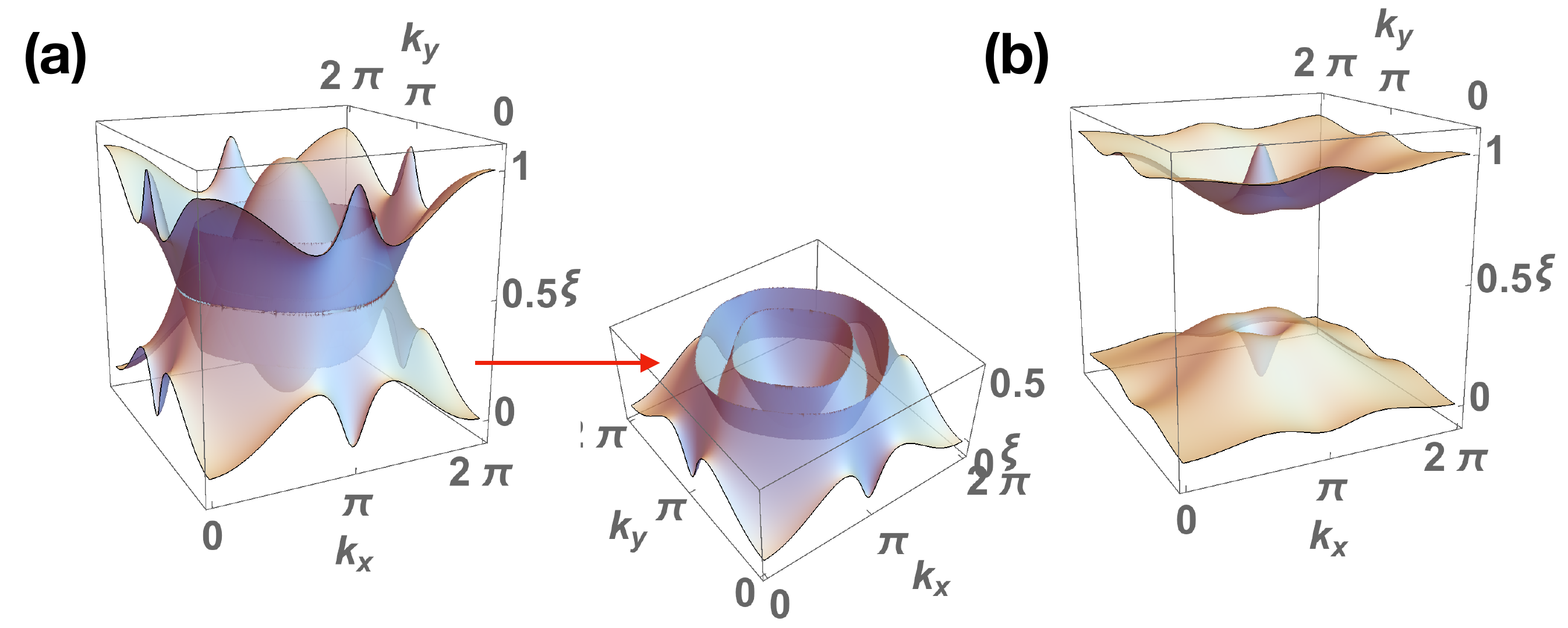}
\caption{Entanglement spectrum  
$\xi(k_x,k_y)$  computed from the post-quench state defined in Eq. (\ref{Eq:13})
for frequency space bipartition with (a) $(t_1, M)=(1,0.5)$, (b) $(t_1, M)=(1,2.5)$.
The right panel in (a) shows the crossings of between the levels are two rings.}
 \label{ee2}
\end{figure}

Similar as two-band models,
the consequence of non-vanishing Hopf number can lead to the crossings of mid-gap states in the entanglement spectrum. 
However, the number of mid-gap states is not directly related to the Hopf number\cite{Deng2013}.
Unlike the two-band models, the entanglement spectrum of the post-quench states in the spin-1 models for a real space bipartition does not show any crossings.
However, if we consider a frequency space bipartition, 
the entanglement spectrum as a function of $(k_x, k_y)$ has mid-gap states
forming two rings for $|\chi|=1$ and is fully gapped for $\chi=0$ as shown in Fig. \ref{ee2}.
These rings are similar to the boundary Fermi rings in Hopf insulators in Refs.\onlinecite{Liu2017,Moore2008, Deng2013}.

\section{Four-band models in 3+1 dimensions}
\label{Sec:4B}

Let us consider a four-band model in three dimensions with the following form
\begin{align}
\Ham_{\bk}=f_\bk \tau_x +g_\bk \tau_y + n_\bk \tau_z \sigma_x + m_\bk \tau_z \sigma_y,
\label{Eq:4b}
\end{align}
where $\{\sigma_i\}$ and $\{\tau_j\}$ are two sets of Pauli matrices.
There are two two-fold degenerate energies  $E_\bk=\pm \sqrt{f_\bk^2 +g_\bk^2 +n_\bk^2 + m_\bk^2}$. 
The unit vector $\hat{d}_\bk=({f}_\bk, {g}_\bk, {n}_\bk,  {m}_\bk)/|E_\bk|$ characterizes the topology of the static Hamiltonian.
Since this unit vector corresponds to a three-sphere $S^3$, the classification of the static Hamiltonian in three dimensions
is the third Homotopy group $\pi_3(S^3) = \mathbb{Z}$.
The associate topological invariant is the three dimensional winding number
\begin{align}
\nu_3 = \frac{1}{2\pi^2} \int_{\rm BZ} {d}^3k \epsilon^{abcd} \hat{d}_a \partial_{k_x} \hat{d}_b \partial_{k_y} \hat{d}_c \partial_{k_z} \hat{d}_d.
\end{align}

The post-quench state is evolved by the evolution operator $U_\bk(t) =e^{-i \Ham_\bk t} = \cos (|E_{\bf k}| t) - i \mathcal{H}_{\bf k} \sin (|E_{\bf k}| t)$.
It will recur to the initial state at $t=2\pi/|E_{\bf k}|$. 
For simplicity, we normalize the Hamiltonian $|E_\bk|=1$ in the following discussion.

To have non-trivial topology of the post-quench order parameter in $3+1$ dimensions,
the manifold of the post-quench order parameter can be a four sphere $S^4$ such that $\pi_4(S^4)=\mathbb{Z}$.
We consider the initial state $|\psi_i \rangle=(1,0,0,0)^{\rm T}$.
The corresponding post-quench state is
\begin{align}
| \psi_{\bf k}(t) \rangle = \left(\begin{array}{c}\cos t \\-i \sin t (n_\bk + i m_\bk) \\-i \sin t (f_\bk + i g_\bk) \\0\end{array}\right)
\label{Eq:18}
\end{align}
and a post-quench order parameter can be defined as
\begin{align}
{\bf L}&= \langle \psi_{\bf k} (t) | (\tau_x,\tau_y,\tau_z\sigma_x,\tau_z\sigma_y,\tau_z\sigma_z) | \psi_{\bf k}(t) \rangle \notag\\
&= (\hat{g}_\bk \sin 2t, -\hat{f}_\bk \sin 2t, \hat{n}_\bk \sin 2t, - \hat{m}_\bk \sin 2t, \cos 2t),
\end{align}
such that $|{\bf L}|=1$. I.e., the manifold for this post-quench order parameter is a four-sphere $S^4$.
The topology of the post-quench order parameter can be indexed by the second Chern number
\begin{align}
C_2&= \frac{-3}{8\pi^2} \int^{\pi/2}_0 dt \int_{\rm BZ} d^3k  \epsilon^{abcde} L_a \partial_{k_x} L_b \partial_{k_y} L_c \partial_{k_z} L_d\partial_t L_e \notag\\
&=\frac{3}{4 \pi^2} \int^{\pi/2}_0 \sin^3 (2t) \int_{\rm BZ} d^3 k \epsilon^{abcd} \hat{d}_a \partial_{k_x} \hat{d}_b \partial_{k_y} \hat{d}_c \partial_{k_z} \hat{d}_d \notag\\
&=\nu_3.
\end{align} 
Here we demonstrate that the second Chern number of the post-quench order parameter is equal
to the three-dimensional winding number of the static Hamiltonian.
Notice that the period of the post-quench order parameter is $\pi$
and there is an intrinsic symmetry,
\begin{align}
&L_a(t) \partial_{k_x} L_b(t) \partial_{k_y} L_c(t) \partial_{k_z} L_d(t)\partial_t L_e(t) \notag\\
=&L_a(-t) \partial_{k_x} L_b(-t) \partial_{k_y} L_c(-t) \partial_{k_z} L_d(-t)\partial_t L_e(-t).
\end{align}
To have non-vanishing second Chern number, the time integral is taking from $0$ to $\pi/2$.

To demonstrate the relation between non-vanishing second Chern number and the entanglement spectrum of the post-quench state, we consider
the functions in the static Hamiltonian in Eq. (\ref{Eq:4b}) being
\begin{align}
f_\bk&=\sin k_x, \quad g_\bk=\sin k_y, \quad n_\bk=\sin k_z, \notag \\
m_\bk&=M-\cos k_x - \cos k_y - \cos k_z.
\label{4bp}
\end{align}

The three-dimensional winding number characterizing  the static Hamiltonian is
\begin{align}
\nu_3 =\left\{
\begin{array}{rl}
&1, \quad 1<|M|<3,  \notag\\
&-2 ,  \quad |M|<1, \notag\\
&0, \quad 3<|M|.
\end{array}
\right.
\end{align}

\begin{figure}[!htbp]
\center
\includegraphics[width=\columnwidth] {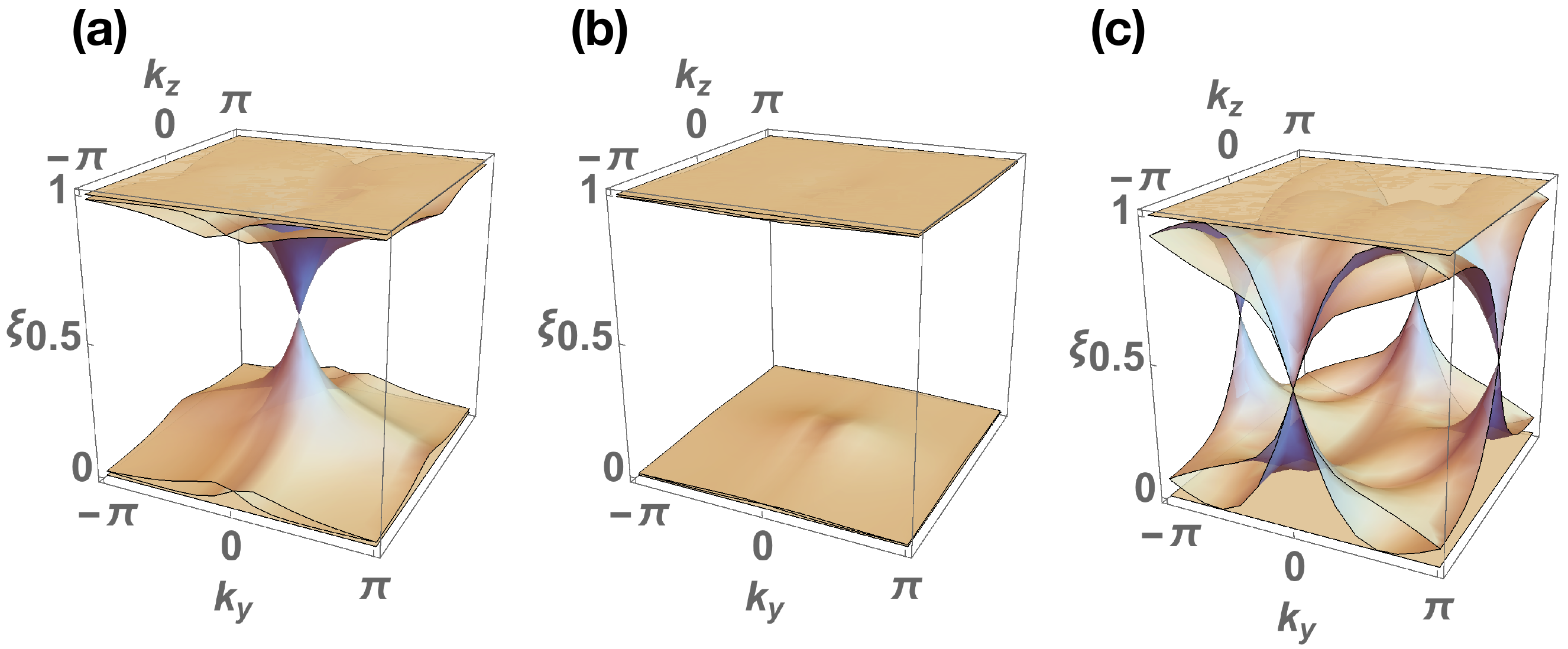}
\caption{Entanglement spectrum $\xi(k_y, k_z)$  computed from the post-quench state defined in Eq. (\ref{Eq:18})
for a real space bipartition at $t=\pi/2$ with (a) $M=2$, (b) $M=4$, and 
(c) $M=0$, where $M$ is the parameter in Eq. (\ref{4bp}).}
 \label{ee3}
\end{figure} 

The entanglement spectrum of the post-quench state reflects the non-trivial topology of the post-quench state. 
The Berry connection associated with the post-quench states
are
\begin{align} 
&A_t = 0, \notag\\
&A_{k_i} = - \sin^2 t (n\partial_{k_i} m - n\partial_{k_i} m+f\partial_{k_i} g-g\partial_{k_i} f ).
\end{align}
We can compute the three dimensional polarization from the Chern-Simons 3-form of the the post-quench state
\begin{align} 
P_3&= \frac{-1}{4 \pi^2} \int_{\rm BZ}  d^3 k\epsilon^{\alpha \beta \gamma} A_\alpha \partial_{\beta} A_\gamma  \notag\\
&=\frac{1}{2 \pi^2} \sin^4 t \int_{\rm BZ}  d^3 k \epsilon^{abcd} \hat{d}_a \partial_{k_x} \hat{d}_b \partial_{k_y} \hat{d}_c \partial_{k_z} \hat{d}_d \notag\\
&=   \nu_3 \sin^4 t  .
\end{align}

From the point of view of the dimensional reduction\cite{Qi2008,Ryu2010},
the second Chern number can be computed by the three-dimensional polarization
\begin{align} 
C_2 = \int_{0}^{\pi/2} dt \partial_t P_3(t)  = \nu_3.
\end{align}
At $t=\pi/2$, the post-quench state describes a three-dimensional topological
insulator with the three dimensional polarization $P_3=C_2$.
Hence the non-vanishing second Chern number characterizing 
the post-quench order parameter leads
to Dirac-cone like excitations in the entanglement spectrum of the post-quench state for a real space bipartition.
In the case that $C_2=-1$, there is one Dirac cone at $(k_y, k_z, t) = (0,0,\pi/2)$ [Fig. \ref{ee3} (a)]. 
In the case that $C_2=-2$, there are two Dirac cone at  $(k_y, k_z, t) = (0,\pi,\pi/2)$ and $(\pi,0,\pi/2)$  [Fig. \ref{ee3} (c)]. 
And the entanglement spectrum is fully gapped for $C_2=0$  [Fig. \ref{ee3} (b)].
The number of Dirac cones in the entanglement spectrum is equal to the second Chern number characterizing 
the post-quench order parameter.
On the other hand, the Berry connection $A_t = 0$ indicates no mid-gap states in the entanglement Hamiltonian
for a frequency space bipartition. Numerically, we observe  the entanglement spectrum are fully gapped for either trivial (vanishing second Chern number) or
 topological (non-vanishing second Chern number) post-quench states.


\section{Discussion}
\label{Sec:CD}

\begin{table}
  \centering
\begin{tabular} { | c || c | c | c | }
  \hline
                                & $d=1$ & $d=2$ & $d=3$ \\
   \hline 
  $\pi_d(\mathcal{M}_{\Ham_\bk})$           & $\pi_1(S^1)\to \nu$& $\pi_2(S^2)\to C$ & $\pi_3(S^3)\to \nu_3$ \\  
   \hline
  $\pi_{d+1}(\mathcal{M}_{\hat{n}_\bk(t)})$      & $\pi_2(S^2)\to C_{\rm dyn.}$& $\pi_3(S^2) \to \chi$ & $\pi_4(S^4)\to C_2$ \\
  \hline
\end{tabular}
    \caption{$\mathcal{M}_{\Ham_\bk}$ is the manifold of the static Hamiltonian and $\mathcal{M}_{\hat{n}_\bk (t)}$ is the manifold of
the post-quench order parameter. The classification based on the homotopy groups gives the topological invariants of the static Hamiltonian in $d$ dimensions
and the post-quench order parameter in $d+1$ dimensions. These two topological invariants are related.}
  \label{TAB}
\end{table}

We demonstrate the relation between the topological invariant of the static Hamiltonian in $d$ dimensions
and the topological invariant of the post-quench order parameter in $d+1$ dimensions by use of homotopy groups.
We show that the entanglement spectrum
of the post-quench state reveals its topological property.
If the post-quench order parameter has non-vanishing topological invariant, the entanglement spectrum has mid-gap states 
forming Dirac cones or rings.
Our results are summarized in Table \ref{TAB}.

The thriving developments of cold-atom experiments provide
a way to measure the dynamics of the post-quench states by the method
of Bloch state tomography\cite{Alba2011,Hauke2014,Flaschner1091}.
For example, in a hexagonal optical lattice,  one can prepare a localized cold-atom cloud only at one of the sub-lattices ($A$ sites)  
and let it evolve by a sudden quench to a Chern insulator.
The non-vanishing dynamical Chern number will lead to momentum-time skyrmion
which can be mapped out by the momentum-time resolved  Bloch state tomography.
This measurement in principle can be extended to spin-1 models and the four-band models
with non-trivial topology in cold-atom systems\cite{Cooper2018}.

Before we close the discussion, we would like to point out some future directions.

\begin{itemize}

  \item It has been proposed theoretically that a measurement protocol to access the entanglement spectrum 
  can be realized in cold atoms experiments\cite{Pichler2016}. The mid-gap states in the entanglement spectrum in principle
  can be observed in our quench setup.
  
  \item  Up to date, the entanglement property in the frequency space is only discussed 
  in a two-photon state\cite{Law2000,Kues2017,Jen2016}. 
  It will be an intriguing task for finding the experimental realization 
in condensed matter and cold-atom systems.

  \item Although time direction is period for the post-quench state,
  time is NOT exchangeable with the momentum. This no-exchangeability reflects the property of
  the entanglement spectrum with different bipartitions in real and frequency spaces.
  To understand the condition of the existence of mid-gap states in the entanglement spectrum,
  it is desired to study the structure of the reduced density matrix of the post-quench state for
  different bipartitions in real and frequency spaces.

  \item The interaction and disorder effects in the quench dynamics are an interesting direction to study. 
           In one-dimensional case, it is shown that the disorder does not remove the crossings in the entanglement spectrum under symmetry constraints\cite{Gong2017}.
           It will be interested to investigate the robustness of mid-gap states in the entanglement spectrum for higher dimensional cases.

  \item One possible extension of our analysis is to consider the higher order homotopy group. With the development of the synthetic dimensions\cite{Boada2012,Celi2014,Price2017},
           our method can generalize to dimensions higher than $3+1$. One interesting model\cite{PYC} is a six-dimensional four-band model [Eq. (\ref{Eq:4b})]
           with a post-quench order-parameter on $S^4$. The static Hamiltonian can be classified by $\pi_6(S^3)=\mathbb{Z}_{12}$ and the post-quench order-parameter 
           can be categorized by 
         $\pi_{6+1}(S^4)=\mathbb{Z} \times\mathbb{Z}_{12}$ described by the
           second Hopf fibration $S^7 \to S^4$.
           
\end{itemize}

\begin{acknowledgments}
The author would like to thank Natan Andrei, Piers Coleman, Elio K\"onig, Yashar Komijani, Xueda Wen, Bi Zhen for valuable discussions. 
The author  especially thank Victor Drouin-Touchette for carefully reading the manuscript and helpful comments.
This work was supported by the Rutgers Center for Materials Theory group postdoc grant
(P.-Y. C.).
\end{acknowledgments}

\bibliographystyle{apsrev4-1}
\bibliography{references}

\begin{thebibliography}{52}
\bibitem [{FB()}]{FB}%
  \BibitemOpen
  \href@noop {} {\bibinfo  {journal} {Quantum states can be entangled in the
  frequency domain. One can compute its entanglement property from the
  frequency space bipartition}\ }\BibitemShut {NoStop}%
\end{thebibliography}

\end{document}